# Persistence of Physics and Engineering Students via Peer Mentoring, Active Learning, and Intentional Advising

K. McCavit[1*] and N. E. B. Zellner[2]

[1]Department of Math and Computer Science, 611 E. Porter St., Albion College, Albion, MI USA 49224 (kmccavit@albion.edu), * To whom correspondence should be addressed

[2]Department of Physics, 611 E. Porter St., Albion College, Albion, MI USA 49224

## Abstract

Albion College, a private, undergraduate-only, liberal arts college in Michigan, USA, has developed and implemented a low-cost peer-mentoring program that blends personal and academic support to help students achieve academic success in the introductory courses required for the Physics Major or the Dual-Degree Program in Engineering. This enhanced mentoring program provides much–needed assistance for undergraduate students to master introductory physics and mathematics coursework, to normalize the struggle of learning hard material, and to accept their identity as physics or engineering students (among other goals). Importantly, this program has increased retention among entering science, technology, engineering and mathematics students at Albion College as they move through the introductory classes, as shown by a 20% increase in retention from first-semester to third-semester physics courses compared to years when this program was not in place.

# I. INTRODUCTION

Throughout the United States, a significant percentage of students who enter college with an interest in a science, technology, engineering, or mathematics (STEM) discipline decides to move to a different academic area, often after just the first semester [1]. Research further suggests that first-year STEM students who enroll in less challenging math courses, who take lighter loads, or whose performance in STEM courses is lower relative to non-STEM courses are more likely to switch out of STEM fields [2]. Indeed, there has been ample research over the last decade describing the retreat of students from STEM disciplines during the first and second years of college, and much of this research points to students' lack of proficiency with mathematics [2,3]. This insufficient mathematics preparation of incoming undergraduates presents a serious issue for those charged with preparing the next generation of physicists and engineers.

A typical first-year college course load for students interested in studying physics or engineering would include calculus, but a large proportion of students entering college in the United States do not meet the requirements for entrance into college-level mathematics. In 2013, only 44% of graduating high school students taking the ACT exam  met the college readiness benchmark for mathematics set forth by ACT; of 2014 graduates, only 43% met this benchmark, and this percentage has remained between 42% and 46% since 2009 [4,5]. Thus many prospective physics or engineering students may have to overcome the obstacle of mathematics remediation before enrolling in even the first physics course in the required introductory sequence. Those students who do arrive on campus with sufficient mathematics background to allow immediate enrollment in calculus still face challenges, such as adapting to the level of rigor, adjusting to the amount of work and the pace of the course, and learning how to apply knowledge and skills learned in calculus to topics presented in physics or engineering courses [3]. Therefore, any program targeting retention of students in these areas must include a component that supports development of students' mathematics skills, as well as perseverance, during first-year courses that can be extremely demanding.

At Albion College, students' initial placement into mathematics courses is decided largely by a department-designed placement examination. In Fall 2013, just 49% of prospective physics and engineering students met enrollment requirements for first semester calculus; for such students entering in 2014 and 2015, just 46% and 51%, respectively, were able to enroll directly in calculus. This presents a real issue for students as the physics or pre-engineering curriculum is usually rigorous and stacked, ideally starting in the first year with the calculus-based physics sequence. Therefore, in order to support the student in achieving his or her academic goals, including completing a program on-time, it is imperative that he or she be prepared for academic success in each of the physics and mathematics classes. Many students, however, are already behind math-wise before the first semester even starts. Moreover, while there is interest from both men and women in majoring in STEM fields, few actually graduate with a STEM degree [6,7,], and the numbers of female [9,10] and minority [9,11] STEM graduates are much lower compared to those of majority male STEM graduates.

In response to these challenges, members of the Albion College Department of Physics, the Department of Mathematics and Computer Science, and the Quantitative Studies Center have been exploring strategies to recruit and retain students who desire careers in Physics or Engineering. Mentoring is a proven technique to retain STEM students, but Albion College has developed a blended support system that interweaves proven teaching strategies, intentional advising, and peer mentoring. We show that our students' success is measured by increased satisfaction with Physics or Engineering as a career choice.

## II.     PROGRAM AT ALBION COLLEGE

Albion faculty and staff have worked together to develop a program that focuses on evidence-based teaching techniques already used in the classroom and newly implemented intensive advising and peer mentoring, all of which have the combined potential to impact the academic trajectories of these students. Importantly, this program is relatively low-cost (~$1500 per semester), supported by grader, tutor, and teaching assistant funds that are allocated for such tasks in the annual department budgets of the Department of Physics and the Quantitative Studies Center. This program was implemented on a pilot basis during the 2013-2014 and 2014-2015 academic years, and is continuing in the 2015-16 academic year. Enrollment and retention in all of the Physics courses have increased since the program was implemented.

### A. Teaching practices

Introductory physics and calculus classes at Albion are small in size, typically averaging about 25 students in each section. Research has shown that small classes allow for increased interaction between faculty and students, as well as among students, and that active learning techniques are well-supported and prevalent in the small-class environment [12]. Faculty and staff have used these strategies in order to maximize the percentage of students initially declaring an interest in the Physics Major or Minor or the Dual Degree Program in Engineering (DDPE) who persists to graduation and completes their program on-time. Active learning strategies have also been shown to increase student engagement and satisfaction [13], and Albion faculty routinely employ such activities in the introductory physics and calculus courses. Strategies that are currently in use in these courses include reading quizzes, share/pair, cooperative learning, board work, flipped classrooms, and technology-based activities. Moreover, such interactive-engagement methods have been shown to be more effective than the traditional lecture approach in promoting the learning of physics concepts [14,15], and may contribute to the success of women in physics and engineering [16].

## B. Advising

Activities related to recruiting and retaining students with STEM interests begin during student orientation in the summer, when students who express an interest in physics or engineering are identified and contacted individually by faculty in the Department of Physics and/or the Department of Mathematics and Computer Science. Common practice at Albion's student orientation is to randomly assign students to a faculty advisor; instead the targeted students are personally and individually advised by a faculty member with specific knowledge of the requirements for completion of the STEM program of interest. In particular, students are advised of the series of mathematics courses that must be completed as pre- or co-requisites to required physics courses. Importantly, any student whose mathematics placement is below first-semester calculus is advised with a plan for completing the required series of mathematics courses so that study in physics may commence.

After this initial advising, faculty and staff in the program work together to provide follow-up contact and advising with the goal of showing each student a path to program completion, regardless of the point of entry into mathematics. Because completing the list of required courses, pre-requisites and co-requisites in a timely manner necessitates a highly structured approach to planning the student schedule, advising plays a critical role for these students [17,18]. Thus, by pairing an intentional early intervention at orientation together with continuing contact throughout the fall semester, students are provided essential information and support. Furthermore, by contacting students again after final grades are posted for the fall semester, any unforeseen issues (such as the need to retake a course) are addressed so that students may continue to make progress. Though time- and labor-intensive, this strategy helps both the student and the faculty realize the potential for student success.

### C. Peer mentoring

Research indicates that mentoring shows promise in student retention [19,20]. We have thus added this component to the use of proven teaching practices in the classroom and intentional advising. Mentoring programs may be implemented in various ways, with some utilizing faculty mentors and others incorporating peer mentors, but studies show that periodic meetings with more advanced students in a friendly and supportive environment provide first-year STEM students with valuable encouragement and support as they learn how to be successful college students in general, and STEM students in particular [20,21].

Peer mentors at Albion College are carefully chosen from students who have completed at least the first two years of physics and calculus courses. Through an application and interview process, students are selected to be peer mentors based on mastery of the course material, appropriate personality (outgoing, approachable, patient), and willingness to help other students succeed. Assigned by a staff member, students are put into peer groups (normally between five and eight students per group) according to mathematics background and enrollment in mathematics or physics courses, so that groups are as academically homogeneous as possible. In this way, the peer mentor is able to work with his or her group members on either physics or mathematics material. Peer group mentors are responsible for contacting their group members, setting up meetings (usually two-to-three hours per week, more if an exam is scheduled), keeping in contact with the physics and mathematics professors, and reporting meeting summaries to faculty and staff (via a shared Google document). While some meetings are informal social gatherings, others are more academic sessions focused on study or review of material. To facilitate study sessions, mentors are provided access to student assignments and practice materials, including solutions when available. Historically, students in the target group have been able to find support from faculty during office hours, or from peer tutors on campus. While these options remain available for participants in the mentoring program, students are now intentionally linked to the select subset of peer tutors who have been identified as mentors. Students are thus encouraged to meet with the identified mentors, but they are also able to seek support from other peer tutors or faculty. Importantly, Albion College is an undergraduate-only institution and therefore does not employ graduate students; all of Albion's peer tutors and mentors are undergraduate students who are paid an hourly wage for their work. Mentors take advantage of no-cost activities such as free movies at the local theater, but may incorporate other activities that are not free. For example, one mentor group gathered for ice cream sundaes; another ordered pizzas for a midterm exam review session. We have not seen direct costs to mentors other than this type of food expense, and  costs incurred by the mentors have been reimbursed by the department within two weeks.

Mentors act as tutors for the physics course and also for the mathematics course when applicable, but an essential component of the program is that mentors also act as coaches for students to help students understand the expectations in physics and mathematics classes (e.g., using office hours and study tables, the amount of homework, the pace, effective study strategies, etc.). Mentors are also able to assist students in navigating the expectations of the Physics Major and the DDPE. Finally, mentor meetings give students an outlet for discussing the struggle of adapting to college-level work, and mentors can help students see that this time of transition is a normal part of the first year of college. It has been shown that the ability to successfully navigate the transition to the new norms and expectations of college leads to student success [22]; therefore, a program that facilitates this transition is providing valuable support. An important component of this program is that there is two-way communication between the mentors and the faculty and staff. For example, mentors report to faculty and staff if they have concerns about any student who is struggling with the material or transition to college.

## III.    INCREASING RETENTION AND SUPPORTING SUCCESS

The combination of intensive advising and peer-mentoring, added to proven teaching practices in the classrooms, appears to have increased retention among Physics and DDPE students at Albion College. Enrollment in Analytical Physics I in both Fall 2013 and Fall 2014 showed a large number of first-year students who intended to have careers in Physics, Astrophysics, or Engineering, and the majority of them registered for subsequent Physics and Mathematics classes in both their second and third semesters (see below). Additionally, intensive advising played an important role: students from the entering class of Fall 2013 and 2014 who were at risk of leaving the program because they were not mathematically prepared for Analytical Physics I were recaptured into introductory classes in the subsequent academic year. These students had expressed an interest in physics or engineering during the summer orientation session and were advised to take the appropriate mathematics course(s) during their first year. As communicated to academic advisors and Physics and DDPE faculty and staff during advising sessions, they were keen on continuing the Physics or DDPE program of study. Thus, after successfully completing the prerequisite mathematics course(s) by the end of their first year, they enrolled in Analytical Physics I and Calculus I in their sophomore year.

## IV.    PROGRAM EVALUATION

The pilot peer-mentoring program was evaluated in two ways. First, the percentage of students who self-identified as potential physics or DDPE students moving on from Analytical Physics I in the Fall of each year to Analytical Physics II in the Spring was calculated and compared to previous years. At Albion, it is important that students complete these courses in at least the first two semesters in order to stay on track for timely completion of all required courses. If a student does not complete the physics sequence in consecutive semesters, evidence shows that the student is at risk of not completing the program or not completing the program on-time [2]. The percentage of students moving on from Analytical Physics I to Analytical Physics II was over 90% in both 2013-14 and 2014-15. Prior to the implementation of the peer-mentoring program, the aggregate retention rate for this transition point, using data from 2008-09 through 2012-13, was about 65%. Notably, the retention from first-semester physics through to the third semester (Phys 245: Electronics and Phys 243: Mathematical Methods in Physics I) also increased, with an aggregate 50% retention rate from 2008-09 through 2012-13 compared to an aggregate 71% retention rate from 2013-14 through 2014-15; this indicates that very few students were lost over the summer from the major during those years in which the peer mentoring program has been in place. Figure 1 depicts these retention data.  Any effect on mathematics enrollment is more difficult to quantify, due primarily to the lack of a single path through the required mathematics courses. However, due to the structure of pre-requisite and co-requisite mathematics courses in the Physics Major and DDPE, a student must progress through the required mathematics curriculum in order to enroll in the corresponding physics courses. Therefore, an increase in mathematics performance and retention is an inherent corollary of an increase in retention of physics and engineering students.

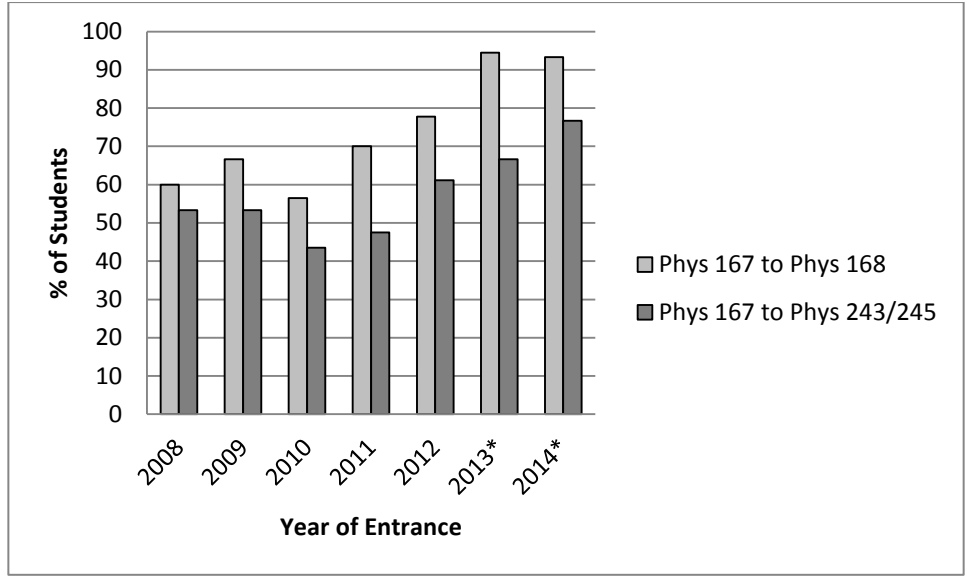

**Figure 1.** Trends in retention, from the first-semester Physics course (Phys 167: Analytical Physics I) to the second-semester Physics course (Phys 168: Analytical Physics II) and to the third-semester Physics courses (Phys 245: Electronics, Phys 243: Mathematical Methods in Physics I), of self-identified Physics Majors and students in the DDPE program (by entering class) since Fall 2008. The * indicates the years that include the peer-mentoring program.

As a second evaluation method, students enrolled in first-semester physics were asked to fill out a survey early in the semester (the pre-test), and a similar survey at the end of the semester (the post-test). The intent of these surveys was to gather information about student attitudes and intentions and to be able to compare the results of the pre-test and post-test to look for patterns or other interesting results. Survey responses for two academic years (Table 1) show that students' attitudes toward physics and mathematics courses at Albion did not appear to deteriorate over the semester, and that, importantly, more than 80% of the respondents were planning to take the next course in the physics sequence. Additionally, at the end of the semester more than half of the respondents could see themselves as future scientists or engineers.

| | Academic Year 2013-14 $n_{pre}$= 26 (87%) $n_{post}$= 26 (87%) | | Academic Year 2014-15 $n_{pre}$= 24 (89%) $n_{post}$= 21 (78%) | |
|---|---|---|---|---|
| **Question** | **PRE-TEST Respondents: "Definitely"** | **POST-TEST Respondents: "Definitely"** | **PRE-TEST Respondents: "Definitely"** | **POST-TEST Respondents: "Definitely"** |
| "I will take a Physics class at Albion next semester." | 62% | 85% | 63% | 81% |
| "I will take a Math class at Albion next semester." | 58% | 65% | 58% | 67% |
| "I will be a mathematician, scientist, or engineer in the future." | 50% | 58% | 63% | 81% |

**Table 1.** Summary of responses to key questions on the pre-semester and post-semester surveys.

## V.     CONCLUSIONS AND FUTURE PLANS

In an effort to retain students in introductory physics and math courses so that they can achieve their goal of becoming physicists or engineers, Albion College faculty and staff have developed and implemented a peer-mentoring program that has spanned two academic years and is continuing. While class instruction includes interactive-engagement techniques, intentional advising by faculty and staff along with peer mentoring by more advanced students appears to further help first-year students to master introductory physics and mathematics coursework, to normalize the struggle of learning hard material, and to accept their identity as physics or engineering students. Due to the apparent success of the pilot program in terms of retention and attitudinal outcomes, a peer-mentor program combined with the use of proven pedagogical techniques in the classroom and intentional advising seems to be a promising model for delivering academic and other important support for entering physics and engineering students at Albion College. Viewing the pilot program as a starting point, an enhanced mentoring program including increased training for mentors, an internship component for students, and external assessment is being planned.

## ACKNOWLEDGEMENTS

The authors thank the following Albion College Physics and Mathematics professors who have worked and continue to work with the peer mentors: Charles Moreau, Aaron Miller, David Seely, and Darren Mason. They also thank the following Albion College peer mentors for participating in the program: Marina Baker, Stefan Blachut, Austin Denha, Jonathan DiNunzio, Alessio Gardi, Alyssa Obert, Matthew Prosniewski, Timothy Szocinski, Robin Todd, Oana Vesa, and Leanne Wegley. The authors also thank Dr. Vanessa McCaffrey and Dr. Melissa Mercer-Tachick for providing valuable comments on an earlier draft of this manuscript and Autumn Bernicky for administrative assistance. We also thank two anonymous reviewers for helpful comments.